\documentclass[12pt,preprint]{aastex}

\shorttitle{WZ Sge}
\shortauthors{Godon et al.} 

\begin{document}

\title{{\it{HST/STIS}} Spectroscopy and Modeling 
of the Long Term Cooling of 
WZ Sagittae following the July 2001 Outburst
\footnote{ Based on observations with
the NASA/ESA {$ Hubble\ Space\ Telescope$}, obtained at the Space
Telescope Science Institute, which is operated by the Association of
Universities for Research in Astronomy, Inc., under NASA contract
NAS5-26555.}}

\author{Patrick Godon\footnote{
Visiting at the Space Telescope Science Institute, 
Baltimore, MD 21218, godon@stsci.edu}, 
and Edward M. Sion}
\affil{Department of Astronomy \& Astrophysics, Villanova
University, Villanova, PA 19085;
patrick.godon@villanova.edu emsion@ast.villanova.edu}

\author{Fuhua Cheng}
\affil{Center for
Astrophysics, University of Science and Technology of China, Hefei,
Anhui 230026, People's Republic of China}

\author{Knox S. Long}
\affil{Space Telescope Science Institute,
3700 San Martin Drive, Baltimore, MD 21218;
long@stsci.edu} 

\author{Boris T. G\"ansicke} 
\affil{Department of Physics, University of
Warwick, Coventry CV4 7AL, United Kingdom; 
Boris.Gaensicke@warwick.ac.uk}

\author{Paula Szkody}
\affil{Department of Astronomy, University of
Washington, Seattle, WA 98195; szkody@astro.washington.edu}

\clearpage 

\begin{abstract}

We present the latest {\it{Hubble Space Telescope (HST) Space Telescope
Imaging Spectrograph (STIS)}} E140M spectrum 
of the dwarf nova WZ Sge, obtained in July 2004, 
3 years following the early superoutburst of July 2001. 
This far-ultraviolet (FUV) spectrum covers the
wavelength interval 1150-1725 \AA , revealing 
Stark-broadened Ly$\alpha$ absorption and absorption
lines due to metals from a range of ionization states.  
The Ly$\alpha$ and C{\small{IV}} double peak emissions  
are still present, indicating the presence of an 
optically thin disk.  
Single white dwarf synthetic spectral fits 
(using $\log{g}=8.5$) to the data
indicate that the white dwarf has now reached  
a temperature $T \approx 15,000 \pm 500$K.

Three years after the outburst the WD is still $\sim$1500K above its
quiescent temperature, it has an FUV flux level almost twice  
its pre-outburst value, and its spectrum does not distinctly exhibit  
the quasi-molecular hydrogen feature around 1400 \AA\ 
which was present in the {\it{IUE}} and 
{\it{HST/GHRS}} pre-outburst data.    
This is a clear indication that even three years after outburst 
the system is still showing the effect of the outburst.  

Taking into account previous temperature estimates obtained 
during the earlier phase of the cooling, 
we model the cooling curve of WZ Sge,
over a period of three years,  
using a stellar evolution code including accretion
and the effects of compressional heating.
Assuming that compressional heating alone is the source of the
energy released during the cooling phase,  
we find that (1) the mass of the white dwarf must be quite
large ($ \approx 1.0 \pm 0.2 M_{\odot}$); and (2) the mass accretion 
rate must have a time-averaged (over 52 days of outburst) value 
of the order of $10^{-8}M_{\odot}$yr$^{-1}$ or above. The
outburst mass accretion rate derived from these
compressional heating models is larger than the rates 
estimated from optical observations \citep{pat02} 
and from a FUV spectral fit \citep{lon03} 
by up to one order of magnitude. This implies that during the cooling
phase the energy released by the WD is not due to 
compressional heating alone. We suggest that ongoing accretion 
during quiescence 
at a moderately low accretion rate can also release a significant amount
of energy in the form of boundary layer irradiation, which
can increase the temperature of the star by several thousand 
degrees.

\end{abstract}


\section{Introduction}

With an orbital period of 82 minutes, 
a recurrence time of 33 years,
and a visual magnitude ``jumping'' from 15 to
8 during outburst, 
the well-studied and widely known system WZ Sge 
is the prototype of a group of H-rich cataclysmic
variables that have the shortest orbital period,  
longest outburst recurrence time
and largest outburst amplitude
of any class of dwarf novae \citep{how99,how02}.
With a distance of only $\approx 43.4$ pc 
\citep{tho03,har04}
it is also the closest cataclysmic variable,  
and the brightest dwarf nova.
The inclination of the binary is high enough
(78 degrees) that the secondary star eclipses the disk rim but not the
white dwarf. 
Recent estimates of the mass of the accreting 
white dwarf range from $\approx 0.8 M_{\odot}$ to $1.2 M_{\odot}$ 
\citep{spr98,ski00,ste01,lon04}.
\citet{ste01} found the radial velocity
semi-amplitude, $K_{2}$, of the secondary
star to be in the range 493 km$~$s$^{-1}$ 
to 585 km$~$s$^{-1}$ 
giving an upper limit to the 
mass of the secondary of $M_2 < 0.11 M_{\odot}$.

The system went into outburst in 1913, 1946, and 1978 
and its recurrence time was therefore assumed to be 33 years.  
However, on 23 July 2001 the system went into
a premature outburst, 10 years earlier than
expected. It was first reported by T. Ohshima 
(see \citet{ish01}), and was then the object of a multi-wavelength
campaign. The July 2001 outburst was the most thoroughly watched
dwarf nova eruption in history.
The system was in outburst for a total of 52 days (consisting of a 
primary burst lasting 24 days, followed by a series of
echo outbursts) and   
on the 53rd day of outburst, the active large accretion phase ended
and the system started to fade without
any other noticeable outburst event 
(for a complete description of the outburst in the optical 
see \citet{pat02}).   

The outbursts of dwarf novae are believed to be based on a 
thermal instability of the accretion disk surrounding the WD.  
During outburst the disk undergoes a transition 
from a low-temperature neutral state
to a high-temperature ionized state. Even though accretion disks appear
in many systems of different sizes (ranging from active galactic nuclei
to low-mass X-ray binaries),  
dwarf novae (because of their observability) 
are still the best laboratories to study the physics of
accretion disks. 
And because of its proximity and intensity, the 2001 July outburst of
WZ Sge presented an ideal opportunity to study the outburst mechanism,
which prompted extensive observations from optical
to X-ray wavelengths. 

{\it{Far Ultraviolet Spectroscopic Explorer (FUSE)}} 
(with a wavelength range of 905-1182 \AA )  
and {\it{HST/STIS}} (covering 1150-1730 \AA )  
spectra of the dwarf nova WZ Sge were obtained
during and following the early superoutburst of July 2001, 
over a time span of 2 years
\citep{kni02,lon03,sio03,lon04}. The system evolved during and
after the primary outburst and by the end of the rebrightening
(echo outbursts) phase the spectrum started to be dominated by the
emission of the cooling WD.  
As the system went into quiescence, the WD temperature 
was determined accurately. The results showed a cooling  
in response to the outburst by about 12,000K, from 
$\approx 28,000$K in 2001 September to $\approx 16,000$K 
in 2003 March \citep{lon03,sio03,lon04}. In a previous work 
\citep{god04} we considered the above FUV observations of WZ Sge 
during the cooling phase, and, using two different methods, 
we independently re-derived the temperature of the white dwarf.  
In this manner we accurately assessed the error bars 
of the cooling curve of the white dwarf. 
In that work, we then modeled the heating and
subsequent cooling of the white dwarf 
using a one-dimensional quasi-static evolution code,   
and found that the mass of the WD must be large ($\approx 1.2 M_{\odot}$) 
and it must accrete at a high accretion rate ($ 9 \times 10^{-9}
M_{\odot}$yr$^{-1}$) if compressional heating is the only source of
energy released by the WD during quiescence. We concluded that ongoing 
accretion at a low rate during quiescence is needed to explain  
the observations if the WD has a mass $M=0.9 M_{\odot}$ and is
accreting at a rate $ \dot{M} = 3 \times 10^{-9} M_{\odot}$yr$^{-1}$
during outburst as inferred by the FUV spectral fits. 

A 28-29 s periodic oscillation was also 
observed both in the optical \citep{pat02} and in the FUV
\citep{wel97,wel03}
as well as a possible harmonic at 15 s during the 2001 outburst
\citep{kni02}. Since this oscillation is not always present
and does not always have the same period, it is difficult to associate
it with the spin of the WD. In addition, if the 29 s signal was indeed
caused by the rotation of WD, that would imply a rotational velocity
of about 3200km$~$s$^{-1}$  (assuming a 0.9 $M_{\odot}$ WD),
much larger than observed.
The suggestion that it could be ZZ Ceti-like
pulsations of the WD has also been ruled out because the oscillation
period did not change significantly while the WD cooled by many
thousands of degrees \citep{wel03}.
Some of the alternatives left for the origin of
this oscillation could be the inner disk, the boundary (or spread) layer
or even a fast rotating accretion belt. A time analysis of the
July 2004 data will be carried out elsewhere.  

In the current work we present and analyze the latest 
{\it{HST/STIS}} spectra obtained in July 2004, namely,  
3 years after outburst.  This latest
observation provides us with an additional data point in the cooling
curve to assess the long term cooling of the WD.
To be consistent with our previous analysis in \citet{god04}, 
we analyze here the spectrum using exactly the same method.  
We confirm our previous result that the accreting white dwarf
must be fairly massive, as suggested by other types of analyses,  
and must have been accreting during the super-outburst at a
time-averaged mass rate of the order of 
$\dot{M} \simeq 10^{-8} M_{\odot}$yr$^{-1}$.
However, our analysis also requires on-going accretion during the 
cooling phase.

\section{The July 2004 {\it{STIS}} Spectrum: Observations and Description}

The observations took place on 2004, July 11 (starting UT 06:23:35 ; 
1083 days post-outburst),
with {\it{HST/STIS}} using the FUV MAMA detector
configuration in TIME-TAG (photon-counting) 
mode with the medium resolution E140M echelle grating and the 0.2" x 0.2"
aperture. 
In this configuration the wavelength coverage is 1140 - 1735 \AA ,  
centered at 1425 \AA. 
The observations consisted of 5 {\it{HST}} orbits 
totaling 13,700s of good exposure time. 
The primary goal of obtaining 5 consecutive orbits was to determine
the mass of the white dwarf \citep{ste05}.
The data reduction
was carried out with the standard STScI pipeline reduction system, 
namely, using CalSTIS version 2.15c (January 29, 2004).
The spectra from the individual spectral orders of the echelle
grating were spliced and binned to a resolution of 0.1 \AA\  
for our analysis. 
The results of the line identifications are presented in Figure 1
in which we have co-added the 5 individual spectra obtained from 
each {\it{HST}} orbit.  

First, we notice that there is an additional  
drop in the continuum flux level 
(measured between 1425 \AA\ and 1525 \AA ) 
from $\approx 6 \times 10^{-14}$ ergs$~$cm$^{-2}$s$^{-1}$\AA$^{-1}$ 
in March 2003 
to $\approx 4.5 \times 10^{-14}$ ergs$~$cm$^{-2}$s$^{-1}$\AA$^{-1}$ 
in July 2004 (see Table 1 where we recapitulate the {\it{STIS}}
observations starting 2001 September 11), 
suggesting that the white dwarf keeps on cooling.
The 2004 July flux level 
is still 1.8 times larger than the pre-outburst flux level 
recorded in late quiescence by {\it{IUE}} (10-14 years since outburst)
and {\it{HST/GHRS}} (17 years since outburst). 
In Table 2 we list some quiescent {\it{IUE}} spectra
of WZ Sge, the average flux level  
(also measured between 1425 \AA\ and 1525 \AA ) 
is $ \approx 2.6 \times 10^{-14}$ ergs$~$cm$^{-2}$s$^{-1}$\AA$^{-1}$. 
In Figure 2 we draw the  
2004 July 11 {\it{HST/STIS}} spectrum together with the 
1989 August 26 {\it{IUE}} spectra (SWP36885) for comparison.  
Both from Table 2 and Figure 2 it is obvious that WZ Sge has not
yet reached its deep quiescence flux level. 

The spectral coverage of {\it{STIS}} for the E140M echelle grating
setup is fixed (1140 \AA\ - 1725 \AA\ , though the region between
1140 \AA\ and 1150 \AA\ is usually too noisy and is being dropped). 
Consequently all the {\it{STIS}} spectra of WZ Sge considered 
in the present work have all the same spectral coverage. This makes
the suite of {\it{STIS}} observations of WZ Sge all directly
comparable. Therefore, in Table 1 we also list the flux
integrated ($ \int d\lambda$) over the entire spectral range of
STIS, to the power 1/4, for all the epochs, which is proportional to
the effective temperature. In the last column of the Table 
we list the temperatures derived in \citet{lon04} and
the ones re-estimated by \citet{god04} together with 
the present results (presented in Table 4, see next section).  

The most prominent line feature in the spectrum is the very broad
Ly$\alpha$ absorption which we attribute to the high-gravity white dwarf
photosphere. 
One does not see distinclty the H$_{2}$ quasi-molecular absorption
feature (centered around 1400\AA ) in the spectrum, which  
was observed and modeled during deep quiescence in the {\it{HST/FOS}}
spectra of the system \citep{Sionetal1995}. 
This feature is expected to
emerge now as the WD temperature is well below 20,000K. 
The spectral signature of the 
Hydrogen quasi-molecular absorption is a pronounced drop in
flux just shortward of $\approx$1400 \AA , as can be clearly
seen in the {\it{IUE}} spectrum in Figure 2.  

Other prominent features in the spectrum are   
the broad emission wings at C {\small{IV}} (1548 \AA , 1550 \AA , 
flanking deep absorption) and the double-peaked Ly$\alpha$ feature
in emission.  They both might be associated, at least in part, with the 
system and are generally assumed to arise from the disk.  
A close examination of the individual exposures 
reveals that the double-peaked features 
have the blue emission peak higher than the red peak
(this is much more pronounced in the C {\small{IV}}   
than in the Ly$\alpha$).  
A rough measurement of the peak-to-peak separation gives  
5\AA.  At Ly$\alpha$, 1\AA\ corresponds to about 245
km$~$s$^{-1}$ (and the 0.1 \AA\ resolution corresponds to 
25 km$~$s$^{-1}$), 
so that this separation corresponds to roughly $1225 \pm 25$ km/s. 
This leads to a value of the disk velocity $V_{disk}\times \sin{i} 
= 613$ km$~$s$^{-1}$.
For comparison, \citet{ski00} measure $V_{disk} \times \sin{i}
=723$ km$~$s$^{-1}$ from the Balmer H$\alpha$ for WZ Sge in quiescence. 
These values for Ly$\alpha$ suggest a disk-formed line, 
and are consistent with the work of \citet{mas00} who report 
several disk velocities.
The Ly$\alpha$ and C{\small{IV}} double peak emissions  
indicate the presence of an optically thin disk.  
This optically thin disk could contribute to the continuum, 
however, the flux in the core of the Ly alpha line drops close to
zero, therefore limiting the amount of optically thin emission
(or any other thermal low gravity emission). 

The other features are predominantly absorption lines 
covering a broad range of ionization. Of immediate
interest is the mix of ions, the excitation/ionization states of the
transitions and the differences one sees between this spectrum and the
spectra of WZ Sge obtained previously, e.g. Figures 1-4 (for the
September, October, November and December 2001 spectra respectively)
in \citet{sio03} and Figures 3, 2 \& 4 (for the April 2002, August 2002
and March 2003 spectra respectively) in \citet{lon04}. See also
Figure 1 in \citet{lon03} for a direct comparison of the flux level   
of the {\it{HST/STIS}} spectra between September 2001 and March 2003. 
Carbon and Silicon absorption lines are observed for a whole range of
ionization levels: C {\small{I}}, C {\small{II}}, C {\small{III}}, 
C {\small{IV}}, Si {\small{II}}, Si {\small{III}}, and Si {\small{IV}}.  
We also observed Al {\small{II}}, Fe {\small{II}},
O {\small{I}}, N {\small{I}} and N {\small{V}}.  
And we tentatively identify 
S {\small{I}}, S {\small{II}}, S {\small{III}},
Al {\small{I}}, N {\small{II}} , N {\small{III}} 
and possibly also Ni {\small{II}}, Cu {\small{II}}, and Co {\small{II}}.
Many of these absorption lines are due to material
local to WZ Sge, since the close distance of WZ Sge (43pc)
precludes an ISM origin.
We recapitulate all the lines we identified in Table 3. For each line we  
list the line center, the line width, the line shift and
in the last column we specify whether these measurments were carried
out for the combined data (summed over the 5 orbits) or for 
an individual orbit. 

We note that to produce substantial Si {\small{IV}} absorption
lines in the WD photosphere requires temperatures above 
about 25,000K, basically 10,000K larger than the WD 
present temperature. 
This suggests that the Si {\small{IV}} lines do not originate
in the WD photosphere.  
The same is true for N {\small{V}} which requires a much higher 
temperature ($T \sim 80,000$K) than Si {\small{IV}}. 
It is noteworthy here that 
N {\small{V}} is seen in absorption in the dwarf nova 
U Gem during quiescence but does
not share the same velocity as the gravitationally-redshifted white dwarf
photosphere \citep{sio98,lon99}. This feature is thought to arise in an
extended hot region of gas near the white dwarf in U Gem. 

We note that the following strong absorption lines were not 
(or only weakly) seen in 
the 2003 March {\it{HST/STIS}} spectrum: 
C {\small{I}} 1277.3, 1277.5 \AA , 
C {\small{I}} 1328.8-1329.6 \AA , 
and N {\small{I}} 1492.63, 1492.82, 1494.68 \AA , while
N {\small{V}} has become less prominent.  
In the present spectrum the width of all the broad absorption lines 
varies slightly from (HST) orbit to orbit (by up to 0.5 \AA)
and is similar to the width observed in the 2003 March spectrum
\citep{lon04}.
Therefore, within this 0.5 \AA\ accuracy 
the width of the lines is similar to the width measured in 
March 2003 by \citet{lon04}. 

There are also unidentified features between 1430\AA\ and 1440\AA\
which we tentatively identify as a
blend of mainly C {\small{I}} with probably Si {\small{I}} 
and possibly Al {\small{I}}. 
Such 
features were seen in the quiescence spectra
of WZ Sge \citep{Sionetal1995,che97}.  
The C {\small{II}} 1335\AA\ ionization lines 
do not resemble expected photospheric features as they are  
too deep (see also the results in next section).

In order to differentiate between the lines that are associated with
the white dwarf (i.e. with the same velocity and/or broadening) 
and the lines that are not associated with it 
(i.e. associated with the binary but in a shell or external ring), 
we check how the profile of a given absorption line
changes from one (spacecraft) orbit to another. 
The line widths and wavelengths were measured and checked
by a careful inspection of the spectrum. 
We find that the following 
absorption lines have the same width and the same wavelength
during the entire observations: 
C {\small{I}} 1266.42 \AA, 
Al {\small{I}} 1271.77 \AA, 
C {\small{I}} 1277.3, 1277.5 \AA, 
O {\small{I}} 1302.17 \AA, 
N {\small{I}} 1316.04 \AA, 
Cu {\small{II}} 1442.14 \AA, 
Co {\small{II}} 1443.84 \AA, 
and Co {\small{II}} 1456.27 \AA. 
These are very sharp absorption lines and are obviously not
associated with the WD nor with its companion  as they do not exhibit any
velocity change: 
within the accuracy of our binning of 0.1 \AA\ all these lines are
at the expected laboratory wavelength.  
Because of the proximity of WZ Sge we identify these lines
with the system albeit outside the binary itself. 
On the other hand, the broader and deeper 
absorption lines exhibit a wavelength
shift of about 0.3-0.5 \AA\ from (spacecraft) orbit to orbit,
corresponding to a velocity of up to
120km$~$s$^{-1}$, where
in the first orbit the line appears to be blue-shifted and 
in a later orbit (usually the 4th) the line appears to be at
its laboratory wavelength. 
However, for the N {\small{V}} doublet lines,
it seems to be the opposite: in the first orbit the lines are at the rest
frame velocity and in the later orbits the line are red-shifted
by 0.4 \AA . This is more clearly observed for the 
N {\small{V}} 1242.80 \AA\ line than for the  
N {\small{V}} 1238.82 \AA\ line, as  
the N {\small{IV}} doublet lines are not very strong and the
flux level at this wavelength is rather low. 
If correct, this points to the fact that the N {\small{V}}
lines might form 
in a region which has a velocity of $\approx 100$ km$~$s$^{-1}$ 
(away from the observer) relatively to the velocity of the WD,
similar to what is observed for U Gem. 
The N V lines in U Gem have velocities which are different  from the
other lines both in outburst and quiescence \citep{sio98,lon99}.

A detailed study of the radial velocities (RV study) of the absorption lines
together with a K1 determination is being carried out elsewhere
\citep{ste05}.

\subsection{The Synthetic Spectral Modeling of the July 2004 Spectrum of WZ Sge} 

In order to determine the parameters of the white dwarf from
the {\it{HST/STIS}} spectrum, we compare the observed
spectrum with a grid of theoretical synthetic spectra generated
with Ivan Hubeny's model atmosphere and spectrum synthesis codes 
TLUSTY200 and SYNSPEC48 \citep{hub88,hub94,hub95}.  We take the
white dwarf photospheric temperature T$_{eff}$, its gravity 
$\log{g}$, its photospheric chemical abundances, and its 
rotational velocity $V_{rot} \sin{i}$ as free parameters.
The comparison is then carried out using a $\chi^2_{\nu}$ minimization 
fitting procedure \citep{numrec}. More details about the minimization
technique, including the scaling parameter, can also be found in
\citet{sio03}.  

In preparation for the fitting of the models we mask 
these regions of the
spectrum which exhibit spectral features not
originating in the WD photosphere and which can signficantly affect
the fitting:   
the N {\small{V}} doublet in absorption around 1240 \AA ; 
the Si {\small{IV}} doublet in absorption around 1400 \AA ;
and both the Ly$\alpha$ (around 1215 \AA ) and C{\small{IV}}
(around 1550 \AA ) double peak emissions. 
It is important to note that 
the N {\small{V}} region overlaps the longward wing of the 
Ly$\alpha$ absorption profile, 
which is very sensitive to the temperature.  
The sharp absorption lines which are most probably not associated
with the WD are not masked, as they are so sharp that they do not
affect the fitting at all. In addition some of these sharp absorption
lines are on top of broader absorption lines/features which themselves
could be due to the WD.  

We use three slightly different approaches in parallel
to assess quantitatively the uncertainty in the spectral fitting
technique.  Following the nomenclature we adopted in \citet{god04} 
we denote the three different methods by (a), (b) and
(c), where (c) is really the method used by \citet{lon04}
while (a) and (b) were used in \citet{god04}. 

In (a) each individual spectrum (for each spacecraft orbit) 
is fit to a synthetic stellar atmosphere spectrum, 
Si and C abundances are fitted separately and 
the regions that are masked are: 
1209-1223 \AA , 1236-1246 \AA , 1380-1410 \AA\ and 1538-1560 \AA. 
In (b) and (c) the spectral fit is carried out on the 
co-added spectra (from the 5 orbits) and scaled solar abundances 
are used.  In (c) the masked regions are:  
1206-1226 \AA , 1233-1246 \AA , 1390-1410 \AA\ and 1538-1560 \AA . 
In (a), however, only the region around the Ly alpha core
is cut out but Gaussian emission components are included for
N {\small{V}}, C {\small{IV}} and He {\small{II}} (1640 \AA). 
Method (a) and (b) assumed $\log{g}=8.5$, while in method (c) 
three different values of 
$\log{g}$ were assumed: 8.0, 8.5 and 9.0.  

\subsection{Determination of the WD Parameters} 

The best-fitting white dwarf models to the July 2004  
observations are displayed in Table 4.     
These best-fit results all give fairly reasonable
$\chi^2_{\nu}$ values and, as a result of this, one cannot use the spectral fits 
alone to select the gravity of the WD. 
Instead, one has to use the fact that in the fitting methods, the radius of
the white dwarf is part of the results when scaling the flux to
the distance to the source (43.4 pc).
The radius obtained (as output in Table 4) for the $log(g)=8.5$ models is
consistent with the radius inferred from the $log(g)=8.5$ assumption
(e.g. see the mass radius relation 
from \citet{ham61} or see \citet{woo90} for different 
composition and non-zero temperatures WDs).  
This is not the case for the $log(g)=8.0$ and $log(g)=9.0$ models,
in which the radius is not self-consistent with the $log(g)$ assumption.  

We therefore adopt the $log(g)=8.5$ solution as the
best fit, for which we find that the WD has a temperature 
of $15,000$K$ \pm 500$K (this is the average
temperature of the best fits in Table 3), 
a radius of $6.2 \pm 0.4 \times 10^{8}$cm, 
a rotational velocity between 250 and 700km$~$s$^{-1}$ 
and non-solar abundances.  

For $\log(g)=9.0$ (a $1.2 M_{\odot}$ white dwarf) 
the best fit models lead to temperatures about 1000K larger,
while for $log(g)=8.0$ ($0.6 M_{\odot}$)   
the the temperatures are cooler by about the same amount ($\approx$1000K). 
The rotational velocities and composition, however, 
do not change significantly as a function of $log(g)$. 

As in \citet{god04} we also compute the ``flux-based'' temperature. 
For this purpose we take the 4th root of the 
integrated flux listed in Table 1, and scale the value obtained
with the same factor as in \citet{god04}, such that for the
March 2003 data,  the flux-based temperature fits the temperature 
assessed using method (b). 
The flux-base temperatures fall all within the average values
estimated by method (a), (b) and (c) for the entire 3 year
period of observations.  
The flux level in the first orbit of July 2004 
was slightly larger than in orbits 2-5, 
and this translates as an increase in temperature of about 400K.

In Figure 3 we display the $\log{g}=8.5$ best fit model using method (c), 
in Figure 4 we display the best fit model using method (b), and
in Figure 5 we display the best fit model to orbit 2 using method (a).
All the other best-fit models listed in
Table 4 are similar to the results we show in Figures 3, 4, 5.  
In these Figures we note that the 4 regions we masked (in
green in Figure 3) are, as expected, in significant disagreement with the 
observations. Another region where the models do not fit the
observed spectrum is in the range $\lambda \approx 1330 - 1365$ \AA ,
where the observed C {\small{II}} $\simeq 1335$ \AA\ and 
C {\small{I}} (+ O {\small{I}} + Si {\small{II}}) $\simeq 1355$ \AA\ 
are much deeper than in the models. It is not clear whether the 
observed excess (in comparison to the models) of flux around 1340 \AA\
is a ``wing'' that belongs to the C {\small{I}} line or to the 
C {\small{II}} line. In model (b) this feature has been
fitted with a Gaussian emitting component. 
At short wavelengths there seems to be
a systematic excess emission (especially noticeable 
on both sides of the double peak Ly$\alpha$
emission). This is probably due to the decline in the
effective area of {\it{STIS}} at short wavelengths. 
Calibration files that correct for this problem do not
currently exist for the echelle grating modes. 
This excess emission could also be due to the presence of
a second component, possibly an optically thin quiescent
accretion disk (see section 3.2). 

The rotational velocity, $V_{rot}\sin{i}$
is of the order of 250 km$~$s$^{-1}$
(as derived by method c) and could
be as high as 700 km$~$s$^{-1}$ (as derived by method a),
depending on the metallicity one uses in the models.
This is about the same value as observed in 2003 March and before.
This velocity, if it corresponds to the true
underlying white dwarf is far lower than the
value of 1200 km$~$s$^{-1}$ found by fitting the GHRS data
\citep{che97} during deep quiescence.
\citet{lon04} re-examined the GHRS data and found
an even higher velocity $V_{rot} \sin{i} \approx$2000 km$~$s$^{-1}$,
but they also noted that there is a correlation between
rotational velocity and metallicity, and that the implied
metallicity was also very high. As the metallicity
increases in the models, the rotational velocity also
must increase in order to match the depth of the lines.
Initially the
{\it{STIS}} observations from to 2001 to March 2003 revealed that
the rotational velocity was increasing with time reaching a
maximum in March 2003 \citep{lon04}. 
It was therefore expected that as the WD
was evolving deeper into quiescence its rotation rate would
eventually match the one observed before outburst with GHRS.
The understanding was that with time the WD was revealing its
stellar surface and that by 2004 we would be able to observe
the rotation of the WD with no or little contamination from line-of-sight
absorption. But the 2004 July observation seems to
indicate a lower rotation rate.

\section{The Heating and Long-Term Cooling of the White Dwarf}

In Table 1 we listed the WD temperature range for the 10 {\it{HST/STIS}}
observations that were obtained 
over a period of 3 years following the 2001 July outburst  
\citep{sio03,lon04}.  
\citet{lon03} also estimated the temperature from two  
{\it{FUSE}} observations. 
The {\it{HST/STIS}} derived temperatures  
were independently re-assessed in \citet{god04} using method (a) 
and (b) and the integrated flux as described in the previous section.
These different methods were
considered in \citet{god04} and here 
in order to explicitly assess the error in the temperature. 
In Figure 6 we plot the observed
cooling curve (with symbols as marked in the graph) 
of the WD as it appears from these observations (including
the two early data points obtained with {\it{FUSE}}, \citep{lon03}). 
The main difference with the observed cooling curve from
\citet{god04} is the addition of the last data point, the 2004 July
{\it{HST/STIS}} data. Before we proceed to the description of the
modeling of the heating and cooling of the WD, it is important to
note that the temperatures listed in Table 1 and displayed 
in Figure 6 were all obtained assuming $\log(g)=8.5$, which 
corresponds to a $0.9 M_{\odot}$ white dwarf mass. 
We recall here that for $\log(g)=9.0$ 
the entire cooling curve shifts upward by about 1000K, 
and similarly for $log(g)=8.0$  
the curve shifts downward by about 1000K.  

In order to model the accretional heating and subsequent cooling
of the WD in WZ Sge, 
we used a 1D evolutionary code without hydrodynamics
(quasi-static assumption).
It is an updated version of the quasi-static stellar
evolution code of \citet{sio95} and more details can
be found there.
We carried out numerical simulations by switching on accretion
for the duration of the superoutburst and then shutting it
off to follow the cooling of the white dwarf.
In this way the effects of compressional heating
can be assessed quantitatively.
The matter was assumed to accrete 'softly' with the same
entropy as the white dwarf outer layers. 
It was also assumed that the accretion and 
heating of the white dwarf 
was uniform rather than being restricted to the equatorial region. 
The transfer of angular momentum (by shear mixing) 
into the white dwarf was neglected.

During the actual outburst, as accretion took place at a
high rate, 
the star's photospheric emission was overwhelmed by the emission of 
the hot components (mainly the inner disk),
which made it difficult to assess the exact temperature of the star
and its rotation rate $\Omega_{wd}$.
However, on day 53 the large accretion phase ended.  
By that time the accretion rate had probably
dropped to its quiescence level. Therefore, we modeled the 
superoutburst of WZ Sge by turning on the accretion (at a constant rate) 
for 52 days, after
which it was shut off and the model was evolved for $\sim$3 years
(1200 days). The exact outburst mass accretion
rate of the system is not known, however, 
it seems very likely that initially the
mass accretion rate was very high at the onset of the outburst and
decreased steadily during the plateau phase and then it was
more erratic during the ``echo outbursts '' phase resembling
a succession of normal dwarf nova outbursts.
It has been shown \citep{god02} that compressional heating
is primarily a function of the accreted mass and that the
time dependence of the mass accretion rate has little effect
on the compressional heating after that mass has been
accreted. This justifies the use of a constant mass accretion rate
in our simulations. In addition, at the present time our code
cannot simulate a variable mass accretion event ($\dot{M}
\ne constant$).

\subsection{Compressional Heating and Subsequent Cooling} 

In the first set of simulations we assume that the observed 
elevated temperature of the star is due to the
compressional heating it has endured during the outburst phase {\it{alone}},
and this energy is released slowly during the cooling phase.  
Namely, we neglect the boundary layer irradiation during outburst, 
as it was shown \citep{god02} that the temperature increase due 
to BL irradiation is sustained only during accretion, and 
when the accretion is turned off, the star rapidly radiates away 
the BL energy absorbed in its outermost layer. However, the
temperature increase due to compressional heating takes place deeper
in the layers of the star and it takes many days (years) for
the star to cool.

We ran models with different white dwarf mass, namely $M_{wd}
= 0.8 M_{\odot}$ to $1.2 M_{\odot}$ 
(by increments of $0.1 M_{\odot}$) and varied 
the mass accretion rate in the range 
$ 10^{-9}M_{\odot}$yr$^{-1} - 10^{-8}M_{\odot}$yr$^{-1}$
(by increments of about 10\% of its value). 
We chose the initial WD temperature ranging from 12,000-15,000K
by increment of 500K.
We take into account the $log(g)/T_{wd}$
degeneracy of the models as stated at the beginning of this section, 
namely, the observed cooling curve displayed in Figure 6 shifts upward
by 1000K when considering the $log(g)=9.0$ solutions and
shifts downward by 1000K when considering the $log(g)=8.0$ solutions. 
For each different WD mass (models 1, 2, 3, 4, \& 5 in Table 5) 
we find a mass accretion rate
that accounts for the elevated temperature of the white dwarf  
with a corresponding simulated cooling curve fitting the data points.  
However, as mentioned previously, 
the synthetic spectral fit in this work and in 
\citet{lon04} are all consistent with 
a $M=0.9 M_{\odot}$ accreting white dwarf. This is the reason why
we display   
in Figure 6 the best fit model for a $0.9 M_{\odot}$ WD (model 2 in Table 5) 
accreting at a rate of $2.8 \times 10^{-8} M_{\odot}$yr$^{-1}$. 
All the other best fits listed in Table 5 (with a different WD mass) 
fitted the observed cooling curve as well (except model 6,
which we consider in the next subsection). 
From these simulations, 
the quiescence WD temperature is $\approx 13,500 \pm 1000$K, and
the average mass accretion rate of the outburst models 
(a few $ 10^{-8} M_{\odot}$yr$^{-1}$) 
is larger by about one order of magnitude than the 
value determined from the spectral fits to the observations
during the outburst phase 
($\approx 3 \times 10^{-9} M_{\odot}$yr$^{-1}$ \citep{lon03}).  
It could be that during the outburst the disk actually 
is self-absorbing/self-occulting and therefore the mass 
accretion rate derived from the observations during outburst
might have been under estimated.  

In the last column of the table we also
list the numerically computed excess energy that is radiated by the 
WD during and following the outburst over 1200 days. 
We see that it is around $2-3 \times 10^{39}$ergs and it increases
to $7-8 \times 10^{39}$ergs when we integrate the numerical 
models over a period of 10,000 days (27.4yrs).  
The computed excess energy radiated by the WD here is only a few
percent of the corresponding theoretical accretion energy 
derived from the mass accretion rate given in Table 5. 
This is similar to \citet{gan96} who 
obtained a value of only 1 percent for VW Hyi. 
For comparison, the total outburst energy 
derived in \citet{pat02} (which is partially based on the FUV data) is 
$4.6 \times 10^{40}$ergs over the 24 days main outburst (20 percent
more is radiated over the next 100 days) and is also
about 10 times less than the accretion energy assumed 
in the compressional heating simulations.  
This is the same discrepancy that was mentioned before
in the mass accretion rate and it arises 
because the accretion rate needed to reproduce the cooling curve 
is more than one order of magnitude larger than derived from the
spectral fits during outburst \citep{lon03}.  
In other words, the accreted mass derived from the observations, 
$\Delta M = 4 \times 10^{23}$g(d/43pc)$^2(M_{wd}/M_{\odot})^{-1.8}$ 
\citep{pat02},  
is one order of magnitude smaller than the accreted mass needed in the 
compressional heating simulations ($ 5.7 \times 10^{24}$g)  
to reproduce the observed cooling curve. Due to the extreme mass
ratio, the difference compared to the compressional heating simulations
could even be larger.  
 
This discrepancy can be accounted for if during the cooling phase
the WD actually releases energy that does not originate
from the compressional heating. 
This energy could include rotational kinetic energy released from 
the outer layer of the star which was spun up during the outburst phase
\citep{kip78,spa93,lon93} and/or ongoing accretion 
after the outburst \citep{lon93,god04}. Since it is not known how
much rotational kinetic energy might be stored in the outer layer
of the star and currently no model exists to assess this
energy (such a model would require two-dimensional simulation of the
accretion including a treatment for the outer envelope of the star), 
we consider here only ongoing accretion after
outburst as a source of additional energy release 
during the cooling phase (see next subsection). 

\subsection{Boundary Layer Irradiation During Quiescence} 

We compute here a $0.9 M_{\odot}$ model (model 6 in table 5) 
with a mass accretion rate
of $\dot{M}= 5 \times 10^{-9} M_{\odot}$yr$^{-1}$,
of the same order (though slighlty larger) 
than the mass accretion rate inferred from the 
{\it{FUSE}} observations on day 7 of the outburst phase  
$\dot{M}= 1-3 \times 10^{-9} M_{\odot}$yr$^{-1}$
\citet{lon03}. 
We display this model in Figure 7, from which it is clear that
compressional heating alone cannot account for the observed
cooling curve : its temperature is too low by $\approx$1,000K in July 2004 
and $6-7,000$K in Fall 2001. 

We suggest here that on-going accretion during quiescence can
increase the temperature of this model to match the observed
cooling curve. As was shown in  
\citet{god04}, during low quiescent accretion,  
boundary layer irradiation can substantially
increase the temperature of the WD.
However, in order to be consistent with the observations, 
this accretion rate must be lower than the limit inferred
from the FUV observations.
The strictest limit on this emission is near 1216 
\AA, where very little flux is expected from the WD, and the observed 
flux level was $5 \times 10^{-15}$ergs$~$cm$^{-1}$s$^{-1}$\AA$^{-1}$.
To assess the maximum accretion 
rate allowed, we consider the simulated disk spectra produced by 
\citet{wad98}.  For or a $0.8 M_{\odot}$ WD with a mass accretion rate of 
$3 \times 10^{-11} M_{\odot}$yr$^{-1}$ 
and an inclination of 78 degrees, the Wade and Hubeny 
simulations suggest a flux of 
$1.6 \times 10^{-14}$erg$~$cm$^{-1}$s$^{-1}$\AA$^{-1}$ at 43 pc. 
Thus,  a reasonable upper limit to the mass accretion rate in July 2004 
was $\simeq 10^{-11} M_{\odot}$yr$^{-1}$.  
For the earlier spectra the upper limit to the 
mass accretion rate increases roughly in the same relative proportion 
to  the continuum flux level near Ly$\alpha$ (see Figure 1 in \citet{lon04}).
As a result, the mass accretion rate then could have been as large as 
$ \approx 5 \times 10^{-11} M_{\odot}$yr$^{-1}$ in 2001 September.   
%

To check this hypothesis of ongoing accretion during the cooling
phase, we carry out simulations of boundary
layer irradiation for model 6 (Figure 7) during the cooling phase,
assuming a temperature 14-16,000K and an accretion rate 
$\dot{M} = 10^{-11}-10^{-10}M_{\odot}$yr$^{-1}$. 
We list the results in Table 6. 
The initial WD temperature $T^i_{wd}$ listed in column 5
represents the temperature of the WD of model 6 in Figure 7
(solid line) around $t \sim 500-1000$ days (14,000K) and 
around  $t \sim 100$ days (16,000K).  
A temperature increase of 1,000K to 4,000K is obtained
for a quiescent mass accretion rate for models 6 through 11. 
However, one would
need an accretion of $\approx 10^{-10} M_{\odot}$yr$^{-1}$
in order to increase the temperature by 6,400K (model 12),
to match the discrepancy of model 9 around day 100. 
We therefore suggest that the real picture is probably somewhere
between model 9 (with accretion during the cooling phase)
and model 2 (where all the heating is due to compressional
heating). Spectral fits to the earlier 
{\it{HST}} data \citep{sio03,lon04} were in better agreement
(in a $\chi^2_{\nu}$ speaking sense) with stellar spectra alone 
than with combined disk-stellar spectra, however the combined
disk-stellar spectra could not be ruled out unambiguously. One of the
problems we are facing here is that at such low accretion rates the
quiescent disk is probably optically thin and accurate
spectral models of its continuum do not exist.           

\section{Conclusion}

Our main purpose in analyzing the 2004 July spectrum of WZ Sge 
was to determine the temperature, mass, rotation rate and chemical 
abundances of the accreting white dwarf in the system. The temperature
was needed to assess the cooling of the WD 3 years after the 
outburst, and we found that the WD's temperature is still higher 
than it was before outburst. We confirmed that the mass of the
WD is large and we inferred from the spectral fit a value of 
$0.9 M_{\odot}$ in agreement with \citet{lon04} and \citet{ste05}. 
%
%
However simulations of post-outburst cooling due to 
the effects of of compressional heating of a $0.9 M_{\odot}$ WD as well as 
ongoing quiescent accretion fit the observed cooling only if the mass 
accretion rate is larger than observed.
Compressional heating simulations with a mass accretion rate
as low as the one inferred from the observations during outburst
have a temperature too low by several thousand degrees and one
needs to assume on-going accretion during the cooling phase
in order to reduce the temperature discrepancy. 

From the spectral fit of the lines we also estimated 
the chemical abundances and the rotation velocity of the WD.  
We expected to find a high stellar rotation rate for the WD
(1200km$~$s$^{-1}$, \citet{che97}) but we found a rotation rate  
of only a few 100km$~$s$^{-1}$. 
However, whether the lines belong
to the WD or to material in the line of sight is unclear. 
\citet{lon04} have shown that slab models can fit the data
better and have suggested caution when deriving  results obtained
from the analysis of the absorption lines, 
as one cannot distinguish which lines in WZ Sge  
originate in the WD photosphere and which ones are due to
circumstellar or interstellar absorption. 
What is now needed is self-consistent modeling of the time-resolved
spectra, which attempts to disentangle the photospheric lines from
those in overlying material. Until such an analysis is successfully
completed, it will not be possible to conclusively state the rotation
velocity of the WD in WZ Sge. 

The N {\small{V}} doublet absorption lines in WZ Sge 
originates from matter that does not share the same velocity
shift as all the other lines (as in U Gem) .   
We also know for sure that the C {\small{IV}} and 
Si {\small{IV}} lines are not from the WD, but they are
from the system, and it is also quite clear that the sharp
absorption lines from 
Al$~${\small{I}}, C$~${\small{I}}, O$~${\small{I}}, Cu$~${\small{I}}, 
Co$~${\small{I}}, and even N$~${\small{I}} may not be associated 
with the white dwarf or cool companion. 
It is not certain, however, that the remaining absorption lines
(which are pretty broad) in the spectrum are from the WD,
as their width slightly changes from orbit to orbit (see section 2). 
This change might be explained if the WD is observed through 
absorbing material that is located asymmetrically above the disk
(the asymmetry could be due to the tidal force of the companion
or to matter overshooting the hot spot).  
From the spectral fit (method a) we find that 
the chemical abundance Si and C both continue to decrease
from 0.4 and 3.5 (respectively) in 2003 March to 0.2 and 2.0 
in 2004 July. 
However, we cannot unambiguously
answer the question whether 
the apparent over-abundance of C relative
to Si reflects the actual situation in the photosphere of 
the white dwarf of WZ Sge.   

%
%
The flux-based temperature method gives a temperature of about 13,500K 
for the 1989 IUE spectra \citep{sle99} 
obtained 11 years after the 1978 outburst which 
had a flux level 30-50 percent lower than in the July 2004 STIS 
spectrum, obtained 3 years after the 2001 outburst.
This temperature is also the quiescent temperature found in
the earlier IUE data \citep{szk89,sio90}.  
It seems that three years after outburst the WD is still 
$\simeq$1500K above its
quiescent temperature, it has an FUV flux level almost twice  
its pre-outburst value, and its spectrum does not distinctly exhibit  
the quasi-molecular hydrogen feature around 1400 \AA\ 
which was present in the {\it{IUE}} and 
{\it{HST/GHRS}} pre-outburst data.    
This is a clear indication that even three years after outburst 
the WD in WZ Sge has not yet returned to its deep quiescent state.  

\acknowledgments
We would like to thank the anonymous referee for her/his constructive
criticism and prompt report. PG is thankfull to the Space Telescope
Science Institute for its kind hospitality.

{} 

\clearpage

\begin{deluxetable}{cccccc}
\tablewidth{0pc}
\tablecaption{
STIS Observation Log with Temperature estimates 
}
\tablehead{
Date        & Day into  &  Exposure & 1425-1525 \AA\ Flux                & (Integrated Flux)$^{1/4}$ &   T$^{(1)}$ \\ 
            &  Outburst &  (s)      &(ergs$~$cm$^{-2}$s$^{-1}$\AA$^{-1}$)& ($10^{-3}$)               &   (1000 K) \\ } 
\startdata
2001 Sep 11 & 50        & 2330      & $  3.9 \times 10^{-13} $           &  3.8                      & 31.9-27.0-28.2   \\ 
2001 Oct 10 & 79        & 2330      & $  2.7 \times 10^{-13} $           &  3.3                      & 25.2-23.6-23.4   \\ 
2001 Nov 10 & 110       & 2330      & $  2.1 \times 10^{-13} $           &  3.1                      & 23.7-22.4-22.1   \\ 
2001 Dec 11 & 141       & 2330      & $  1.7 \times 10^{-13} $           &  2.9                      & 22.6-21.6-20.7   \\ 
2002 Apr 15 & 266       & 5150      & $  1.0 \times 10^{-13} $           &  2.5                      & 19.5-18.8-18.1   \\ 
2002 Jun 05 & 317       & 5150      & $  9.1 \times 10^{-14} $           &  2.5                      & 18.8-18.0-17.4   \\ 
2002 Aug 27 & 400       & 5150      & $  8.0 \times 10^{-14} $           &  2.4                      & 17.8-17.4-16.7   \\ 
2002 Nov 01 & 466       & 5150      & $  6.6 \times 10^{-14} $           &  2.3                      & 17.5-17.0-16.3   \\ 
2003 Mar 23 & 608       & 5150      & $  6.0 \times 10^{-14} $           &  2.2                      & 17.2-16.6-15.9   \\ 
2004 Jul 11 & 1083      & 13700     & $  4.5 \times 10^{-14} $           &  2.0                      & 14.6-15.5-15.4   \\ 
\enddata
\tablenotetext{(1)}
{The temperatures were estimated using three different 
fitting techniques: (a), (b) and (c) respectively. 
Methods (a) and (b) were used in \citet{god04}, 
while method (c) is from \citet{lon04}  
(see section 2.1 for details). 
The 2004 July temperature is from this work  
(see Table 4). All these temperatures were estimated assuming
$\log{g}=8.5$ .} 
\end{deluxetable}     

\clearpage

\begin{deluxetable}{ccccc}
\tablewidth{0pc}
\tablecaption{
IUE Observation Log of WZ Sge in deep quiescence 
}
\tablehead{
Spectrum & Date         &   Exposure & 1425-1525 \AA\ Flux             & Time since \\ 
         &              &   (s)      &(ergs$~$cm$^{-2}$s$^{-1}$\AA$^{-1}$)& outburst (yrs) \\ } 
\startdata
SWP33419 &  1988 May 01 &  10,800    & $  2.27 \times 10^{-14} $  & 10 \\ 
SWP36885 &  1989 Aug 26 &  10,800    & $  2.79 \times 10^{-14} $  & 11 \\ 
SWP45103 &  1992 Jul 08 &  16,500    & $  2.73 \times 10^{-14} $  & 14 \\ 
SWP45298 &  1992 Aug 06 &  15,600    & $  2.45 \times 10^{-14} $  & 14 \\ 
SWP45370 &  1992 Aug 18 &  11,760    & $  2.59 \times 10^{-14} $  & 14 \\ 
\enddata
\end{deluxetable}

\clearpage

\begin{deluxetable}{lcccc}
\tablewidth{0pc}
\tablecaption{
Line Identifications and Absorption Line Measurement, July 2004 Spectrum 
}
\tablehead{
Line Identification                & Line Center  & Line Width & Line Shift  & Orbit      \\ 
                                   & (\AA)        & (\AA)      & (\AA)       &             \\ } 
\startdata
S {\small{I}} 1150.82              & 1150.8       &  0.3       &   ...       & Combined   \\ 
Si {\small{II}} 1155.00            & 1154.9       &  0.3       &    ...      & Combined   \\ 
Si {\small{II}} 1155.96            & 1155.8       &  0.3       &    ...      & Combined   \\ 
S {\small{II}}  1163.20            & 1163.3       &  0.4       &    ...      & Combined   \\ 
S {\small{II}}  1166.15, 1166.63   & 1166.4       &  1.7       &    ...      & Combined   \\ 
C {\small{III}} 1174.60-1176.77    & 1175.2       &  2.5       &    ...      & Combined   \\ 
Si {\small{II}} 1190.42            & 1190.5       &  0.6       &    ...      & Combined   \\ 
Si {\small{II}} 1193.29            & 1193.4       &  0.4       &    ...      & Combined   \\ 
Si {\small{II}} 1194.50            & 1194.5       &  0.5       &    ...      & Combined   \\ 
N {\small{I}}   1200.22            & 1200.3       &  1.0       &    ...      & Combined   \\ 
Si {\small{III}} 1206.50           & 1206.5       &  0.6       &    ...      & Combined   \\ 
N {\small{V}}   1238.82            & 1238.8       &  0.4       &    ...      & Combined   \\ 
N {\small{V}}   1242.80            & 1243.0       &  0.6       &    +0.2     & 1          \\ 
                                   & 1242.8       &  0.6       &    0.0      & 2          \\ 
                                   & 1243.3       &  0.6       &    +0.5     & 4          \\ 
Si {\small{II}} 1250.58            & 1250.6       &  0.4       &    ...      & Combined   \\ 
Si {\small{II}} 1253.81            & 1253.7       &  0.6       &    ...      & Combined   \\ 
Si {\small{II}} 1259.92, 1260.42   & 1260.3       &  1.2       &    ...      & Combined   \\ 
Si {\small{II}} 1264.74, 1265.00   & 1264.8       &  1.0       &    ...      & Combined   \\ 
C {\small{I}}   1266.41            & 1266.2       &  0.1       &    -0.2     & 1,2,3,4,5  \\ 
Al {\small{I}}  1271.77            & 1271.8       &  0.1       &    0.0      & 1,2,3,4,5  \\ 
C {\small{I}}   1277.3, 1277.5     & 1277.4       &  1.0$^a$   &     0.0     & 1,2,3,4,5  \\ 
Si {\small{III}} 1298.89, 1298.95  & 1298.7       &  0.6$^a$   &    -0.1     & 1          \\ 
                                   & 1299.1       &  0.6$^a$   &    +0.3     & 2          \\ 
                                   & 1299.0       &  0.6$^a$   &    +0.2     & 3,5        \\ 
                                   & 1298.8       &  0.6$^a$   &     0.0     & 4          \\ 
O {\small{I}} 1302.17              & 1302.1       &  0.7$^a$   &    -0.1     & Combined   \\ 
Si {\small{II}} 1304.37, 
O {\small{I}} 1304.86              & 1304.6       &  1.1$^a$   &    ...      & Combined   \\ 
O {\small{I}} 1306.03              & 1305.8       &  0.8       &    -0.2     & Combined   \\ 
Si {\small{II}} 1309.28            & 1309.2       &  0.7$^a$   &    -0.1     & Combined   \\ 
N {\small{I}} 1316.04              & 1316.0       &  0.7$^a$   &     0.0     & Combined   \\ 
N {\small{I}} 1319.67              & 1319.7       &  0.2       &     0.0     & Combined   \\ 
S {\small{III}} 1328.16-1328.81    & 1328.7       &  0.3       &     ...     & Combined   \\ 
C {\small{I}} 1328.8               & 1329.1       &  0.3       &     ...     & Combined   \\ 
C {\small{I}} 1329.60              & 1329.6       &  0.2       &     ...     & Combined   \\ 
C {\small{II}} 1334.53, 1335.7     & 1334.7       &  2.4       &    -0.4     & 1          \\ 
                                   & 1334.95      &  2.2       &    -0.15    & 2          \\ 
                                   & 1335.2       &  2.4       &    +0.1     & 3,4,5      \\ 
Si {\small{IV}} 1393.76            & 1393.3       &  1.0       &    -0.4     & 1          \\ 
                                   & 1393.65      &  1.0       &    -0.1     & 2          \\ 
                                   & 1393.85      &  1.0       &    +0.1     & 3,4,5      \\ 
Si {\small{IV}} 1402.77            & 1402.35      &  1.0       &    -0.4     & 1          \\ 
                                   & 1402.85      &  1.0       &    +0.1     & 3,4,5      \\ 
N {\small{III}} 1410.08            & 1410.1       &  0.1       &     0.0     & Combined   \\ 
Si {\small{III}} 1417.24,  
Cu {\small{III}} 1418.43           & 1417.6       &  1.5       &     ...     & Combined   \\ 
C {\small{III}} 1428.18            & 1428.2       &  0.3       &     ...     & Combined   \\ 
Cu {\small{II}} 1442.14            & 1442.2       &  0.1       &     0.0     & 1,2,3,4,5  \\ 
Co {\small{II}} 1443.84            & 1443.9       &  0.1       &     0.0     & 1,2,3,4,5  \\ 
Co {\small{II}} 1456.27            & 1456.5       &  0.1       &     +0.2    & 1,2,3,4,5  \\ 
C {\small{I}} 1463.34              & 1463.2       &  0.2       &     -0.1    & Combined   \\ 
Ni {\small{II}} 1467.26            & 1467.3       &  0.1       &      0.0    & Combined   \\ 
Ni {\small{II}} 1467.74            & 1467.9       &  0.1       &      0.1    & Combined   \\ 
N {\small{I}} 1492.63, 1492.82     & 1492.4       &  0.5       &     -0.3    & 1          \\ 
                                   & 1492.8       &  0.5       &     +0.1    & 4          \\ 
N {\small{I}} 1494.68              & 1494.7       &  0.2$^a$   &      0.0    & 1          \\ 
                                   & 1494.7       &  0.4$^a$   &      0.0    & 4          \\ 
Si {\small{II}} 1526.71            & 1526.5       &  1.0$^a$   &     -0.2    & 1          \\ 
                                   & 1526.8       &  1.0$^a$   &      0.1    & 4          \\ 
Si {\small{II}} 1533.43            & 1533.1       &  1.0$^a$   &     -0.3    & 1          \\ 
                                   & 1533.4       &  1.0$^a$   &      0.0    & 4          \\ 
C {\small{IV}} 1548.20             & 1547.6       &  1.0$^a$   &     -0.6    & 1          \\ 
                                   & 1548.2       &  1.5$^a$   &      0.0    & 4          \\ 
C {\small{IV}} 1550.78             & 1550.8       &  1.5$^a$   &      0.0    & Combined   \\ 
C {\small{I}} 1560.31-1561.44      & 1561.0       &  1.8$^a$   &      ...    & Combined   \\ 
N {\small{II}} 1596.43             & 1596.7       &  0.1       &     +0.3    & Combined   \\ 
Fe {\small{II}} + 
C {\small{I}} 1608.4               & 1608.3       &  0.5       &      ...    & Combined   \\ 
C {\small{I}} 1656.27-1658.12      & 1657.3       &  2.5       &      ...    & Combined   \\ 
Al {\small{II}} 1670.79            & 1670.5       &  0.5$^a$   &      -0.3   & 1          \\ 
                                   & 1670.8       &  0.5$^a$   &      0.0    & 4          \\ 
Fe {\small{II}} 1673.46            & 1673.2       &  0.1       &     -0.3    & 1          \\ 
                                   & 1673.5       &  0.1       &      0.0    & 4          \\ 
S {\small{III}} 1713.11            & 1713.0       &  0.1       &     -0.1    & Combined   \\ 
\enddata
\tablenotetext{a}
{These lines also reveal a sharp absorption line at the bottom of the
broad absorption line.}  
\end{deluxetable}

\clearpage

\begin{deluxetable}{cccccccc}
\tablewidth{0pc}
\tablecaption{
2004 July STIS temperature estimates 
}
\tablehead{
Exposure &  Method & Log(g) &  $T_{wd}$ & $V _{rot}\sin{i} $ & Log R  & Metallicity   & $\chi^2_{\nu}$ \\  
         &         &        &  (1000K)  &  (km/s)      &  (cm)  &  (solar)      &       \\ }
\startdata
Orbit 1  &  (a)    & 8.5    &  14.80    & 500          & 8.816  &  C=2.0 Si=0.2 & 1.113 \\ 
Orbit 2  &  (a)    & 8.5    &  14.60    & 700          & 8.819  &  C=2.0 Si=0.2 & 1.005 \\ 
Orbit 3  &  (a)    & 8.5    &  14.60    & 500          & 8.813  &  C=2.0 Si=0.2 & 0.932 \\ 
Orbit 4  &  (a)    & 8.5    &  14.60    & 600          & 8.813  &  C=2.0 Si=0.2 & 0.900 \\ 
Orbit 5  &  (a)    & 8.5    &  14.70    & 400          & 8.808  &  C=2.0 Si=0.2 & 1.408 \\ 
Combined &  (b)    & 8.5    &  15.50    & 300          & 8.760  &  0.5          & 0.900  \\ 
Combined &  (c)    & 8.0    &  14.44    & 267          & 8.836  &  0.7          & 1.174  \\ 
Combined &  (c)    & 8.5    &  15.41    & 260          & 8.760  &  0.7          & 1.127 \\ 
Combined &  (c)    & 9.0    &  16.45    & 245          & 8.686  &  0.7          & 1.083 \\ 
Orbit 1  & flux    & 8.5    &  15.40    & -            &   -    &     -         &  -  \\ 
Orbit 2-5& flux    & 8.5    &  15.00    & -            &   -    &     -         &  -   \\ 
\enddata
\end{deluxetable}

\clearpage

\begin{deluxetable}{ccccccc}
\tablewidth{0pc}
\tablecaption{
Accretional heating models for WZ Sge
}
\tablehead{
Model  & $M_{wd} $     & $ Log R_{wd} $ & Log(g) & $T_{wd}^{i(1)}$ & $\dot{M}$  & $\Delta E $       \\
number & $(M_{\odot})$ & $(cm)        $ &$(cgs)$ & $(1,000K)$      & $(M_{\odot}/yr)$  & (ergs)  \\
}
\startdata
1 & 0.8 & 8.845   & 8.34 &  12.5   & $4.0 \times 10^{-08}$ & $3.6 \times 10^{39} $       \\
2 & 0.9 & 8.795   & 8.49 &  13.0   & $2.8 \times 10^{-08}$ & $2.9 \times 10^{39} $       \\
3 & 1.0 & 8.745   & 8.64 &  13.5   & $2.0 \times 10^{-08}$ & $2.5 \times 10^{39} $       \\
4 & 1.1 & 8.700   & 8.77 &  14.0   & $1.4 \times 10^{-08}$ & $2.4 \times 10^{39} $       \\
5 & 1.2 & 8.600   & 9.00 &  14.5   & $9.0 \times 10^{-09}$ & $2.3 \times 10^{39} $       \\
6 & 0.9 & 8.795   & 8.49 &  14.0   & $5.0 \times 10^{-09}$ & $7.0 \times 10^{38} $       \\
\enddata
\tablenotetext{1}
{The initial quiescent temperature of the white dwarf at the beginning of each 
simulation.}

\end{deluxetable}

\begin{deluxetable}{rcccccc}
\tablewidth{0pc}
\tablecaption{
Boundary layer irradiation models for WZ Sge in quiescence 
}
\tablehead{
Model  & $M_{wd} $     & $ Log R_{wd} $ & Log(g) & $T_{wd}^{i(1)}$ & $\dot{M}$  & $T_{wd}^{(2)}   $       \\
number & $(M_{\odot})$ & $(cm)        $ &$(cgs)$ & $(1,000K)$      & $(M_{\odot}/yr)$  & $(1,000K)$  \\ 
}
\startdata
7 & 0.9 & 8.795   & 8.49 &  14.0   & $1.0 \times 10^{-11}$ & 15.5      \\
8 & 0.9 & 8.795   & 8.49 &  14.0   & $2.0 \times 10^{-11}$ & 16.6      \\
9 & 0.9 & 8.795   & 8.49 &  14.0   & $3.0 \times 10^{-11}$ & 17.5      \\
10 & 0.9 & 8.795   & 8.49 &  14.0   & $5.0 \times 10^{-11}$ & 19.0      \\
11 & 0.9 & 8.795   & 8.49 &  16.0   & $5.0 \times 10^{-11}$ & 20.0      \\
12 & 0.9 & 8.795   & 8.49 &  16.0   & $1.0 \times 10^{-10}$ & 22.4      \\
\enddata
\tablenotetext{1}
{The initial temperature of the white dwarf at the beginning of each simulation.}
\tablenotetext{2}
{The elevated temperature of the white dwarf due to boundary layer irradiation
assuming a stellar rotation rate $\Omega_*=0.2 \Omega_K$.}

\end{deluxetable}

\clearpage

{\large{\bf{\center{
Figures Caption }}}} 

Figure 1: 
{\it{HST/STIS}} spectrum of WZ Sge obtained on 2004, July 11, 
almost 3 years after optical maximum, with suggested line
identifications. The wavelength is given in Angstr\"om 
(\AA ) and the flux in erg$~$cm$^{-2}$s$^{-1}$\AA$^{-1}$. 
Note the scale is slightly different in each panel. In the
last (bottom) panel, the stars denote regions where the echelles
do not overlap and create gaps.  

Figure 2: 
2004 July 11 {\it{HST/STIS}} spectrum together with the 
1989 August 26 {\it{IUE}} spectra (SWP36885). The flux level
in 2004 (3 years after outburst) 
has still not yet reached the late quiescence (lower) flux level 
observed in 1989 with {\it{IUE}} (11 years after outburst).   

Figure 3: 
July 2004 (combined) spectrum compared with the best-fitting 
log(g)=8.5 model computed in a manner
similar to \citet{lon04} - method (c). Top: Plots of the observed
spectrum (black) and model spectra (red). Data that were excluded
from the fitting (in the mapping technique) are plotted in green.
The excluded regions are the double peak emissions from 
Ly$\alpha$ (H {\small{I}} around 1215 \AA ) 
and C {\small{IV}} (around 1550 \AA ) 
together with absorption lines not originating from the WD atmosphere
(N {\small{V}} 1238.82 \AA , 1242.80 \AA and   
Si {\small{IV}} 1393.76 \AA , 1402.77 \AA ). 
Bottom: Difference between the observed spectrum and the fitted
spectrum (black); statistical error (two blue lines).  

Figure 4: 
July 2004 (combined) spectrum compared with the
best fitting log(g)=8.5 calculated in a manner similar to
\citet{god04} - method (b) (see Table 1 and text for details).  
The top panel shows the observed spectrum in solid black,
the individual model components (WD + Gaussians) in black dotted
lines and the sum of all components as a thick solid gray line.
The bottom panel shows the residual, where the masked region 
(Ly$\alpha$) has been omitted. 

Figure 5: 
July 2004 (exposure \# 2) spectrum compared with the
best fitting log(g)=8.5 calculated in a manner similar to
\citet{god04} - method (a) (see Table 1 and text for details).  

Figure 6: Modeling the heating and cooling of WZ Sge. 
The temperature (in Kelvin) of the
white dwarf is drawn as a function of time (in days) since the start of
the outburst (July 23, 2001).
The solid line represents the compressional 
heating model with a 0.9 solar mass white dwarf (corresponding to
Log(g)=8.5) with an initial temperature 
of 13,000K, accreting at a rate of $2.8 \times 10^{-8} M_{\sun}/$year 
for 52 days.

Figure 7: Same as Figure 6, but here 
the compressional 
heating model has a 0.9 solar mass white dwarf (corresponding to
Log(g)=8.5) with an initial temperature 
of 14,000K, accreting at a rate of $5 \times 10^{-9} M_{\sun}/$year 
for 52 days. The compressional heating alone cannot account for the
observed elevated temperature of the WD.

\clearpage

\begin{figure}
\plotone{f1.eps}     
\end{figure}

\begin{figure}
\plotone{f2.eps}
\end{figure}

\begin{figure}
\plotone{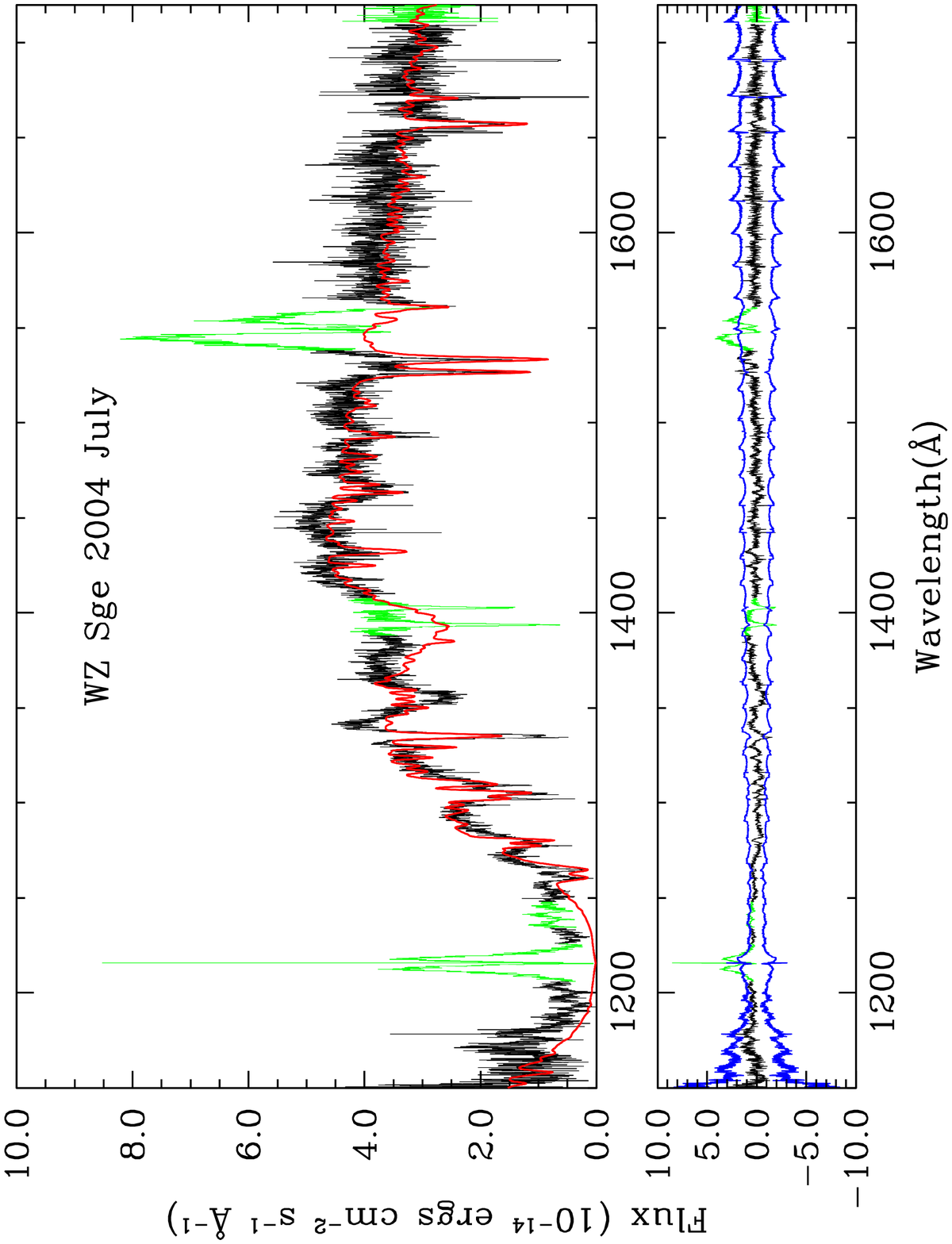}
\end{figure}

\begin{figure}
\plotone{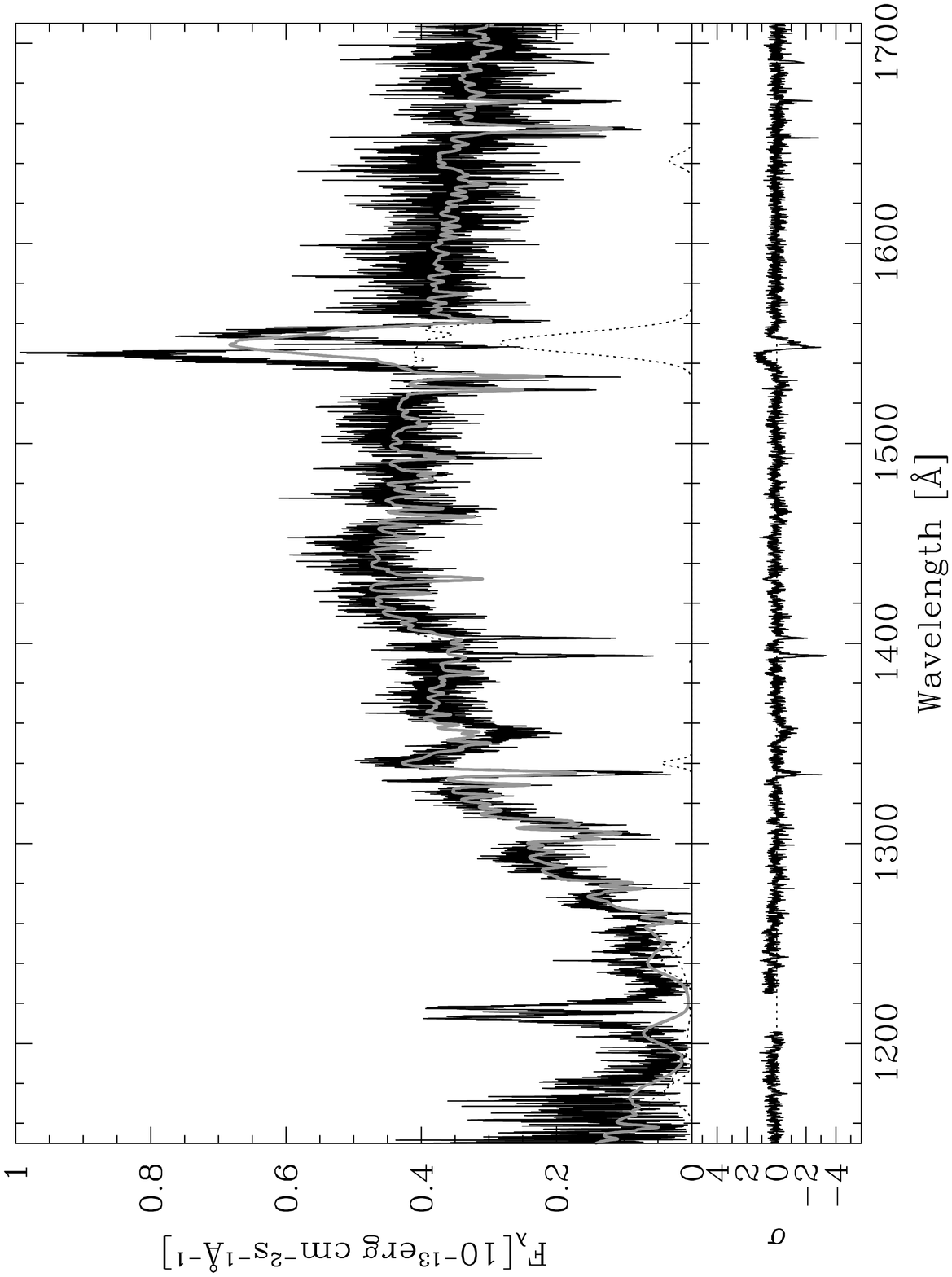}
\end{figure}

\begin{figure}
\plotone{f5.eps}
\end{figure}

\begin{figure}
\plotone{f6.eps}
\end{figure}

\begin{figure}
\plotone{f7.eps}
\end{figure}


\begin{thebibliography}{}

%
\bibitem[Cheng et al.(1997)]{che97}
Cheng, F., Sion, E.M., Szkody, P., \& Huang, M. 1997,
\apj, 484, L149 

\bibitem[G\"ansicke \& Beuermann (1996)]{gan96} 
G\"ansicke, B.T., \& Beuermann 1996, A\&A, 309, L47     

\bibitem[Godon \& Sion (2002)]{god02} 
Godon, P., \& Sion, E. M. 2002, \apj, 566, 1084

\bibitem[Godon et al.(2004)]{god04}
Godon, P., Sion, E.M., Cheng, F., G\"ansicke, B.T., 
Howell, S., Knigge, C., Sparks, W.M., \& Starrfield, S.,
2004, \apj, 602, 336 

\bibitem[Hamada \& Salpeter (1961)]{ham61}
Hamada, T., \& Salpeter, E.E. 1961, \apj, 134, 683  

\bibitem[Hasenkopf \& Eracleous (2002)]{has02}
Hasenkopf, C.A, \& Eracleous, M., 2002, AAS Meeting 201, 120.03 

\bibitem[Harrison et al.(2004)]{har04}
Harrison, T.E., Johnson, J.J., McArthur, B.E., Benedict, G.F.,
Szkody, P., Howell, S.B., \& Gelino, D.M. 2004, \aj, 127, 460 

\bibitem[Howell et al.(1999)]{how99} 
Howell, S.B., Ciardi, D., Szkody, P., van Paradijs, J., Kuulkers, E.,
Cash, J., Sirk, M., \& Long, K.S. 1999, \pasp, 111, 342

\bibitem[Howell et al.(2002)]{how02} 
Howell, S.B., Fried, R., Szkody, P., Sirk, M., 
\& Schmidt, G. 2002, \pasp, 114, 748


\bibitem[Hubeny (1988)]{hub88} 
Hubeny, I. 1988, Comput. Phys. Comm., 52, 103

\bibitem[Hubeny et al.(1994)]{hub94} 
Hubeny, I., Lanz, T.,
\& Jeffrey, S. 1994, Newsletter on Analysis
of Astronomical Spectra (St. Andrews Univ.), 20, 30

\bibitem[Hubeny \& Lanz (1995)]{hub95} 
Hubeny, I., \& Lanz, T. 1995, \apj, 439, 875

\bibitem[Ishioka et al.(2001)]{ish01}
Ishioka, R. et al. 2001, IAU Circ., 7669, 1
%

\bibitem[Knigge et al.(2002)]{kni02} 
Knigge, C., Hynes, R.I., Steeghs, D., Long, K.S., Araujo-Betancor, S.,
\& Marsh, T.R., 2002, \apj, 580, L151 

\bibitem[Knigge et al.(2000)]{kni00} 
Knigge, C., Long, K.S., Hoard, D., Szkody, P., Dhillon, V. 
2000, \apj, 539, L49

\bibitem[Kippenhahn \& Thomas (1978)]{kip78}
Kippenhahn, R., \& Thomas, H.-C. 1978, A\&A, 63, 265 

\bibitem[Kuulkers et al.(2002)]{kuu02} 
Kuulkers, E., Knigge, C., Steeghs, D., Wheatley, P.J., Long, K.S.
2002, in
"The Physics of Cataclysmic Variables and Related Objects", eds. B.T.
G\"ansicke, K. Beuermann, and K. Reinsch, ASP Conf.Ser., 261, 443.

\bibitem[Long \& Gilliland (1999)]{lon99}
Long, K.S., \& Gilliland, R.L. 1999, \apj, 511, 916L 

\bibitem[Long et al.(1993)]{lon93}
Long, K.S., Blair, W.P., Bowers, C.W., Davidsen, A.F., 
Kriss, G.A., Sion, E.M., \& Hubeny, I. 1993, \apj, 405, 327 

\bibitem[Long et al.(2003)]{lon03} 
Long, K.S., Froning, C.S., G\"ansicke, B., Knigge, C., Sion, E.M., 
\& Szkody, P.  2003, ApJ, 591, 1172 

\bibitem[Long et al.(1994)]{lon94}
Long, K.S., Sion, E.M., Huang, M., \& Szkody, P. 1994, \apj, 
424, L49 

\bibitem[Long et al.(2004)]{lon04}
Long, K.S., Sion, E.M., G\"ansicke, B.T., \& Szkody, P., 
2004, \apj, 602, 948 

\bibitem[Mason et al.(2000)]{mas00}
Mason, E., Skidmore, W., Howell, S.B., Ciardi, D.R., 
Littlefair, S., Dhillon, V.S. 2000, MNRAS, 318, 440  
 
\bibitem[Patterson et al.(2002)]{pat02}
Patterson J., et al. 2002, \pasp, 114, 721 

\bibitem[Popham (1999)]{pop99}
Popham, R. 1999, \mnras, 308, 979 

\bibitem[Press et al.(1992)]{numrec}
Press, W.H., Teukolsky, S.A., Vetterling, W.T., Flannery, B.P., 
Numerical Recipes in Fortran 77, The Art of Scientific Computing, 
Second Edition, 1992, Cambridge University Press. 

\bibitem[Regev \& Shara (1989)]{reg89} 
Regev, O., \& Shara, M. M. 1989, \apj, 340, 1006 

\bibitem[Shaviv \& Starrfield (1987)]{sha87} 
Shaviv, G., \& Starrfield, S. 1987, \apj, 321, L51 

\bibitem[Sion (1995)]{sio95} 
Sion, E. M. 1995, \apj, 438, 876 

\bibitem[Sion et al.(1995)]{Sionetal1995} 
Sion, E.M., Cheng, F.H., Long, K.S., Szkody, P., Gilliland, R.L., 
Huang, M., Hubeny, I. 1995, \apj, 439, 957 

\bibitem[Sion et al.(1998)]{sio98}
Sion, E.M., Cheng., F., Szkody, P., Sparks, W.M., G\"ansicke, B.T.,
Huang, M., Mattei, J. 1998, \apj, 496, 449 

\bibitem[Sion et al.(2003)]{sio03} 
Sion, E.M., G\"ansicke, B.T., Long, K.S., Szkody, P., Cheng, F., 
Howell, S.B., Godon, P., Knigge, C., Marsh, T., Sparks, W.M., 
\& Starrfield, S., 2003, \apj, 592, 1137  

\bibitem[Sion \& Szkody (1990)]{sio90}
Sion, E.M., \& Szkody, P. 1990, Proc. IAU Coll. 122, 
``Physics of Classical Novae'', A. Cassatella \& 
R. Viotti (Eds), p.59, Springer-Verlag 

\bibitem[Skidmore et al.(2000)]{ski00} 
Skidmore, W., Mason, E., Howell, S.B., Ciardi, D.R., Littlefair, S.,
\& Dhillon, V.S. 2000, \mnras, 318, 429

\bibitem[Slevinsky et al.(1999)]{sle99}
Slevinsky, R.J., Stys, D., West, S., Sion, E.M., \&
Cheng, R.H. 1999, \pasp, 111, 1292 

\bibitem[Sparks et al.(1993)]{spa93}
Sparks, W.M., Sion, E.M., Starrfield, S., Austin, S. 1993, in 
The Physics of Cataclysmic Variables and Related Objects
(eds. Regev and Shaviv)  

\bibitem[Spruit \& Reuten (1998)]{spr98} 
Spruit, H.C., \& Reuten, R.G.M. 1998, \mnras, 299, 768 

\bibitem[Steeghs et al.(2001)]{ste01} 
Steeghs, D., Marsh, T., Knigge, C., Maxted, P.F.L., Kuulkers, E.,
\& Skidmore, W. 2001, \apj, 562, L145

\bibitem[Steeghs et al.(2005)]{ste05} 
Steeghs, D., Howell, S.B., Knigge, C., G\"ansicke, B.T.,
Sion, E.M. , 2005, in preparation

\bibitem[Szkody \& Sion (1989)]{szk89}
Szkody, P., \& Sion, E.M. 1989, Proc. IAU Coll. 114, 
``White Dwarfs'', G. Wegner (Ed.), p.92, Springer-Verlag  

\bibitem[Thorstensen (2003)]{tho03} 
Thorstensen, J.R. 2003, \aj, 126, 3017 

\bibitem[Wade \& Hubeny (1998)]{wad98} 
Wade, R., \& Hubeny, I. 1998, \apj, 509, 350

\bibitem[Welsh et al.(2003)]{wel03}
Welsh, W.F., Sion, E.M., Godon, P., G\"ansicke, B.T., Knigge, C., 
Long, K.S., \& Szkody, P. 2003, \apj, 599, 509  

\bibitem[Welsh et al.(1997)]{wel97}
Welsh, W.F., Skidmore, W., Wood, J., Cheng, F., \& Sion, E.M.
1997, \mnras, 291, 57   
 
\bibitem[Wood (1990)]{woo90}
Wood, M.A. 1990, Ph.D. thesis, University of Texas at Austin 

\end{thebibliography}
\end{document}